\documentclass[journal=jctcce,manuscript=article]{achemso}
\usepackage[version=3]{mhchem}
\usepackage{caption}
\usepackage{subcaption}
\usepackage{hyperref}
\usepackage{natbib}
\usepackage{siunitx}
\usepackage{tabularx}
\usepackage{amsmath}
\usepackage{amssymb}
\usepackage{xcolor}
\usepackage{threeparttable}
\usepackage{multirow}
\usepackage{booktabs}
\usepackage{subcaption}
\usepackage{braket}

\newcommand{\bfk}{{\mathbf{k}}}

\newcommand{\bfr}{{\mathbf{r}}}
\newcommand{\meV}{\,\mathrm{meV}}
\newcommand{\nsigma}{{n\sigma}}

\author{Han Yang}
\affiliation{Department of Chemistry, University of Chicago, Chicago, Illinois 60637, United States}
\alsoaffiliation{Pritzker School of Molecular Engineering, The University of Chicago, Chicago, Illinois 60637, United States}
\author{Marco Govoni}
\affiliation{Materials Science Division and Center for Molecular Engineering, Argonne National Laboratory, Lemont, Illinois 60439, United States}
\alsoaffiliation{Pritzker School of Molecular Engineering, The University of Chicago, Chicago, Illinois 60637, United States}
\email{mgovoni@anl.gov}
\author{Arpan Kundu}
\affiliation{Pritzker School of Molecular Engineering, The University of Chicago, Chicago, Illinois 60637, United States}
\author{Giulia Galli}
\affiliation{Department of Chemistry, University of Chicago, Chicago, Illinois 60637, United States}
\alsoaffiliation{Pritzker School of Molecular Engineering, The University of Chicago, Chicago, Illinois 60637, United States}
\alsoaffiliation{Materials Science Division and Center for Molecular Engineering, Argonne National Laboratory, Lemont, Illinois 60439, United States}
\email{gagalli@uchicago.edu}

\title{Computational protocol to evaluate electron-phonon interactions within density matrix perturbation theory}
\begin{document}
\maketitle

\begin{abstract}
We present a computational protocol, based on density matrix perturbation theory, to obtain non-adiabatic, frequency-dependent electron-phonon self-energies for
molecules and solids. Our approach enables the evaluation of electron-phonon interaction
using hybrid functionals, for spin-polarized systems, and the computational overhead to include dynamical and non-adiabatic terms in the evaluation of  electron-phonon
self-energies is negligible. We discuss results for molecules, as well as pristine and defective solids.
\end{abstract}

\section{Introduction}\label{sec:intro}
The study of electron-phonon interaction in solids can be traced back to the early days of quantum mechanics\cite{Bloch1929}, and it has been instrumental in  explaining fundamental properties of solids, including  conventional superconductivity\cite{Bardeen1957}. However, it was not until recent years that  electron-phonon interaction was computed from first-principles\cite{Giannozzi1991,Baroni2001,Giustino2007,Giustino2017,Antonius2014}, leading to non-phenomenological  predictions of transport properties of solids\cite{Ponce2018} and of electron-phonon renormalizations of band structures\cite{Antonius2014,Ponce2015,McAvoy2018,Yang2021,Kundu2021}. While early studies relied on semi-empirical models\cite{Fan1950,Fan1951,Froehlich1950} to study electron-phonon interaction, modern investigations typically employ the frozen phonon (FPH) approach,\cite{Monserrat2018,Antonius2014} density functional perturbation theory (DFPT),\cite{Baroni2001,Giustino2007,McAvoy2018,Yang2021,Engel2022} or  molecular dynamics (MD) simulations\cite{Karsai2018,Monserrat2014,Kundu2021}, with electron-electron and electron-ion interactions described at the level of density functional theory (DFT).\cite{Kohn1965}

In two recent papers,\cite{McAvoy2018,Yang2021}  we combined first principles calculations of electron–electron and electron–phonon self-energies in molecules and solids, within the framework of density functional perturbation theory (DFPT). We developed an approach that enables the evaluation of electron–phonon coupling at the $G_0W_0$ level of theory for systems with hundreds of atoms, and the inclusion of non-adiabatic and temperature effects at no additional computational cost. We also computed\cite{Kundu2021} electron-phonon renormalizations of energy gaps by using the path-integral molecular dynamics (PIMD) methods to investigate  anharmonic effects in crystalline and amorphous solids. The DFPT, FPH and MD-based methods are addressing different regimes and different problems; the use of DFPT and FPH are appropriate for systems whose atomic constituents all vibrate {\it close} to their equilibrium positions, although  anharmonic effects have been included in some FPH calculations\cite{Antonius2014}. The assumption of close to equilibrium vibrations is however not required when applying PIMD, which  thus  has a wider applicability; for example it can be used to study amorphous materials, and molecules and solids  exhibiting prominent anharmonic effects, e.g. molecular crystals\cite{Monserrat2015,Kundu2022} and several perovskites\cite{Knoop2020anharmonicity}. However, the calculation of electron-phonon renormalizations using FPH and MD-based methods are carried out within the Allen-Heine-Cardona (AHC)\cite{Allen1976,Allen1981} formalism, which neglects dynamical and non-adiabatic terms of the electron-phonon self-energies. These effects have been shown to be essential to describe electron-phonon interactions in numerous polar materials\cite{Antonius2015,Miglio2020}, for example \ce{SiC}. Perturbation-based methods, on the other hand, can accurately compute electron-phonon self-energies within and beyond the AHC formalism, thus including non-adiabatic and/or frequency-dependent effects into the self-energy. Another benefit of DFPT-based method is the ability to explicitly evaluate the electron-phonon coupling matrices which are useful quantities, for example, in the study of mobilities\cite{Li2015,Ponce2018mobility} and polaron hopping\cite{Sio2019,Sio2019polarons}. 

Here we generalize the perturbation-based approach of Ref.\citenum{McAvoy2018} and \citenum{Yang2021} to enable efficient calculations of electron-phonon interaction with hybrid functionals, by using density matrix perturbation theory  (DMPT)\cite{Waler2006,Rocca2010} to compute phonons and electron-phonon coupling matrices. Our implementation takes advantage of the Lanczos algorithm\cite{Lanczos1950}, which enables the calculations of electron-phonon self-energies beyond the AHC approximation, at no extra cost.

 DMPT has been used in the literature to compute excitation energies and absorption spectra in molecules and solids in conjunction with time-dependent density functional theory (TDDFT)\cite{Rocca2007,Rocca2008}, and to solve the Bethe-Salpeter equation (BSE)\cite{Salpeter1951,Strinati1988,Rocca2010,Rocca2012,Rocca2014,Nguyen2019}. In the latter case,  DMPT has been applied to obtain the variation of single particle wavefunctions due to the perturbation of an electric field. However, DMPT is a general formalism that can be used to compute the response of a system to perturbations of any form, including perturbations caused by atomic displacements.

In this paper, we first derive a formalism for phonon calculations within DMPT, starting from the quantum Liouville equation in \autoref{sec:method}; we then verify our results by comparing them with those of FPH and PIMD calculations in \autoref{sec:verification}. We then present calculations of electron-phonon  interactions in small molecules (\autoref{sec:small-molecule}), pristine (\autoref{sec:diamond}) and defective diamond (\autoref{sec:defect}) using hybrid functionals and we conclude the paper in \autoref{sec:conclusion} with a summary of our findings.
\section{Methodology}\label{sec:method}
Using Hartree atomic units ($\hbar = e = m_e = 1$), we describe the electronic structure of a solid or molecule within Kohn-Sham (KS) density functional theory and we consider the  quantum Liouville's equation to describe perturbations acting on the system:
\begin{equation}
    i\frac{\mathrm{d}}{\mathrm{d}t} \gamma(t) = [H^\mathrm{KS}(t),\gamma(t)],
\end{equation}
where $[\cdot,\cdot]$ denotes a commutator, $H^\mathrm{KS}(t)$ is the Kohn-Sham Hamiltonian
\begin{equation}
    H^\mathrm{KS} = K + V_\mathrm{H} + V_\mathrm{ext} + V_\mathrm{xc},
\end{equation}
with $K$ the kinetic operator, $V_\mathrm{H}$ the Hartree potential, $V_\mathrm{ext}$ the external potential and $V_\mathrm{xc}$ the exchange-correlation potential. The KS Hamiltonian does not depend explicitly on time and depends implicitly on time through the time-dependent density matrix $\gamma$,  that can be written in terms of Kohn-Sham single-particle orbitals $\psi_{n\sigma}$
\begin{equation}
    \gamma(\bfr,\bfr^\prime;t) = \sum_{\sigma}\sum_{n}^{N_\mathrm{occ}^\sigma}\psi_{n\sigma}(\bfr;t)\psi_{n\sigma}^*(\bfr^\prime;t),
\end{equation}
where $\sigma$ is the spin polarization, $n$ is the band index and $N_\mathrm{occ}^\sigma$ is the number of occupied bands in the spin channel $\sigma$. Below we present calculations performed by sampling the Brillouin zone with only the $\Gamma$ point and hence omit labeling eigenstates with $\bfk$-points.

Given a time-dependent perturbation $\partial V_\mathrm{ext}(t)$ acting on the Hamiltonian, the first order change of the density matrix $\partial \gamma (t)$ satisfies the following equation,
\begin{equation}
    i\frac{\mathrm{d}}{\mathrm{d}t} \partial\gamma(t)= \mathcal{L}\cdot\partial\gamma(t)+[\partial V_\mathrm{ext}(t),\gamma_0],\label{equ:td-quantum-liouville}
\end{equation}
where $\mathcal{L}$ is the Liouville super-operator,
\begin{equation}
    \mathcal{L}\cdot\partial\gamma(t) = [H^\mathrm{KS}_0,\partial\gamma(t)]+[\partial V_\mathrm{H}[\partial\gamma(t)],\gamma_0]+[\partial V_\mathrm{xc}[\partial\gamma(t)],\gamma_0].
\end{equation}
Here we use the notation $\partial$ to represent a change of potentials ($\partial V$), wavefunctions $\partial\psi$, charge densities $\partial \rho(\bfr)$ and density matrices $\partial \gamma(\bfr,\bfr^\prime)$; $\gamma_0$ and $H^\mathrm{KS}_0$ are the density matrix and the Kohn-Sham Hamiltonian of the unperturbed system, respectively. 

Taking the Fourier transform of  Eq.~\eqref{equ:td-quantum-liouville}, we rewrite it in the frequency domain,
\begin{equation}
    (\omega-\mathcal{L})\cdot\partial\gamma(\omega) = [\partial V_\mathrm{ext}(\omega),\gamma_0].\label{equ:Liouville-full-freq}
\end{equation}
In phonon calculations, we adopt the Born-Oppenheimer approximation\cite{Born1927} and no retardation effects are included. Hence, we  only need to solve Eq.~\eqref{equ:Liouville-full-freq} at $\omega=0$,
\begin{equation}
    \mathcal{L}\cdot\partial\gamma = -[\partial V_\mathrm{ext},\gamma_0].
\end{equation}
The equation above can be cast in the following form:
\begin{equation}
    \left[
        \begin{array}{cc}
            \mathcal{D}+\mathcal{K}^{1e}-\mathcal{K}^{1d} & \mathcal{K}^{2e}-\mathcal{K}^{2d}              \\
            -\mathcal{K}^{2e}+\mathcal{K}^{2d}            & -\mathcal{D}-\mathcal{K}^{1e}+\mathcal{K}^{1d}
        \end{array}
        \right]
    \left[
        \begin{array}{c}
            \mathcal{A} \\
            \mathcal{B}
        \end{array}
        \right]
    =
    \left[
        \begin{array}{c}
            \left\{\left|-\mathcal{P}^c\partial V_\mathrm{ext}\psi_{n\sigma}\right\rangle:n=1,\cdots,N_\mathrm{occ}^\sigma;\sigma= \uparrow,\,\downarrow \right\} \\
            \left\{\left|\,\,\,\,\mathcal{P}^c\partial V_\mathrm{ext}\psi_{n\sigma}\right\rangle:n=1,\cdots,N_\mathrm{occ}^\sigma; \sigma=\uparrow,\,\downarrow \right\}
        \end{array}
        \right]\label{equ:Liouville}
\end{equation}
where $\mathcal{P}^c$ is the projection operator onto the virtual bands; $\mathcal{K}^{1e}$, $\mathcal{K}^{2e}$, $\mathcal{K}^{1d}$, $\mathcal{K}^{2d}$, defined below in Eq.~\eqref{equ:k1e-lda-gga}--\eqref{equ:k2e-lda-gga} and Eq.~\eqref{equ:k1e-hybrid}--\eqref{equ:k2d-hybrid} are related to the variation of exchange-correlation potential $V_\mathrm{xc}$; the elements of the arrays $\mathcal{A} = \{a_\nsigma:n=1,\cdots,N_\mathrm{occ}^\sigma;\sigma=\uparrow,\downarrow\}$ and $\mathcal{B}=\{b_\nsigma:n=1,\cdots,N_\mathrm{occ}^\sigma;\sigma=\uparrow,\downarrow\}$ are variations of wavefunctions; the variation of the density matrix in terms of wavefunction variation is:
\begin{equation}
    \partial \gamma = \sum_{\sigma}\sum_{n}^{N_\mathrm{occ}^\sigma}\big[\left|a_\nsigma\right\rangle\left\langle\psi_\nsigma\right| + \left|\psi_\nsigma\right\rangle\left\langle b_\nsigma\right|\big]\label{equ:change-of-gamma}
\end{equation}
In phonon calculations, the external perturbation is static $\partial V_\mathrm{ext}(\omega=0)$, and Eq.~\eqref{equ:Liouville} can be further simplified since $a_\nsigma = b_\nsigma = \partial\psi_\nsigma$ for static perturbations and Eq.~\eqref{equ:Liouville} becomes:
\begin{equation}
    \left[\mathcal{D}+\mathcal{K}^{1e}-\mathcal{K}^{1d} + \mathcal{K}^{2e}-\mathcal{K}^{2d} \right] \mathcal{A}
    =
    \left\{\left|-\mathcal{P}^c\partial V_\mathrm{ext}\psi_\nsigma\right\rangle:n=1,\cdots,N_\mathrm{occ}^\sigma;\sigma= \uparrow,\,\downarrow\right\}.\label{equ:general-stern}
\end{equation}

Eq.~\eqref{equ:general-stern} is a generalized Sternheimer equation,\cite{Sternheimer1954} where the operators on the left hand side are defined below.
\begin{equation}
    \mathcal{D}\mathcal{A} = \left\{\mathcal{P}^c(H_0^\mathrm{KS}-\varepsilon_\nsigma)\ket{a_\nsigma}:n = 1,\cdots,N_\mathrm{occ}^\sigma;\sigma= \uparrow,\,\downarrow\right\}.
\end{equation}

When using  LDA/GGA functionals, the $\mathcal{K}^{1e}$ and $\mathcal{K}^{2e}$ operators are:
\begin{equation}
    \mathcal{K}^{1e}\mathcal{A} = \left\{
    \int\mathrm{d}\mathbf{r}^\prime\mathcal{P}^c(\mathbf{r},\mathbf{r}^\prime)\psi_\nsigma(\mathbf{r}^\prime)\sum_{\sigma^\prime}\sum_{n^\prime}^{N_\mathrm{occ}^{\sigma^\prime}}\int\mathrm{d}\mathbf{r}^{\prime\prime}f_\mathrm{Hxc}(\mathbf{r}^\prime,\mathbf{r}^{\prime\prime})\psi_{n^\prime{\sigma^\prime}}^*(\mathbf{r}^{\prime\prime})a_{n^\prime {\sigma^\prime}}(\mathbf{r}^{\prime\prime}):n=1,\cdots,N_\mathrm{occ}^\sigma;\sigma= \uparrow,\,\downarrow
    \right\},\label{equ:k1e-lda-gga}
\end{equation}

\begin{equation}
    \mathcal{K}^{2e}\mathcal{A} = \left\{
    \int\mathrm{d}\mathbf{r}^\prime\mathcal{P}^c(\mathbf{r},\mathbf{r}^\prime)\psi_\nsigma(\mathbf{r}^\prime)\sum_{\sigma^\prime}\sum_{n^\prime}^{N_\mathrm{occ}^{\sigma^\prime}}\int\mathrm{d}\mathbf{r}^{\prime\prime}f_\mathrm{Hxc}(\mathbf{r}^\prime,\mathbf{r}^{\prime\prime})a_{n^\prime {\sigma^\prime}}^*(\mathbf{r}^{\prime\prime})\psi_{n^\prime{\sigma^\prime}}(\mathbf{r}^{\prime\prime}):n=1,\cdots,N_\mathrm{occ}^\sigma;\sigma= \uparrow,\,\downarrow\label{equ:k2e-lda-gga}
    \right\},
\end{equation}
where $f_\mathrm{Hxc} = v_\mathrm{c} + f_\mathrm{xc}$ is the sum of the bare Coulomb potential $v_c$ and the exchange-correlation kernel
\begin{equation}
f_\mathrm{xc}(\bfr,\bfr^\prime) = \frac{\delta V_\mathrm{xc}(\bfr)}{\delta\rho(\bfr^\prime)}
\end{equation}
with $\rho(\bfr)$ being the electron density.

When using hybrid functionals, the operators are:
\begin{equation}
    \mathcal{K}^{1e}\mathcal{A} = \left\{
    \int\mathrm{d}\mathbf{r}^\prime\mathcal{P}^c(\mathbf{r},\mathbf{r}^\prime)\psi_\nsigma(\mathbf{r}^\prime)\sum_{\sigma^\prime}\sum_{n^\prime}^{N_\mathrm{occ}^{\sigma^\prime}}\int\mathrm{d}\mathbf{r}^{\prime\prime}f^\mathrm{loc}_\mathrm{Hxc}(\mathbf{r}^\prime,\mathbf{r}^{\prime\prime})\psi_{n^\prime{\sigma^\prime}}^*(\mathbf{r}^{\prime\prime})a_{n^\prime{\sigma^\prime}}(\mathbf{r}^{\prime\prime}):n=1,\cdots,N_\mathrm{occ}^\sigma;\sigma= \uparrow,\,\downarrow\label{equ:k1e-hybrid}
    \right\},
\end{equation}

\begin{equation}
    \mathcal{K}^{2e}\mathcal{A} = \left\{
    \int\mathrm{d}\mathbf{r}^\prime\mathcal{P}^c(\mathbf{r},\mathbf{r}^\prime)\psi_\nsigma(\mathbf{r}^\prime)\sum_{\sigma^\prime}\sum_{n^\prime}^{N_\mathrm{occ}^{\sigma^\prime}}\int\mathrm{d}\mathbf{r}^{\prime\prime}f^\mathrm{loc}_\mathrm{Hxc}(\mathbf{r}^\prime,\mathbf{r}^{\prime\prime})a_{n^\prime{\sigma^\prime}}^*(\mathbf{r}^{\prime\prime})\psi_{n^\prime{\sigma^\prime}}(\mathbf{r}^{\prime\prime}):n=1,\cdots,N_\mathrm{occ}^\sigma;\sigma= \uparrow,\,\downarrow\label{equ:k2e-hybrid}
    \right\},
\end{equation}

\begin{equation}
    \mathcal{K}^{1d}\mathcal{A} = \left\{
    \alpha\int\mathrm{d}\bfr^\prime\mathcal{P}^c(\bfr,\bfr^\prime)\sum_{n^\prime}^{N_\mathrm{occ}^\sigma}a_{n^\prime\sigma}(\bfr^\prime)\int\mathrm{d}\bfr^{\prime\prime}v_c(\bfr^\prime,\bfr^{\prime\prime})\psi_{n^\prime\sigma}^*(\bfr^{\prime\prime})\psi_{n\sigma}(\bfr^{\prime\prime}):n=1,\cdots,N_\mathrm{occ}^\sigma;\sigma= \uparrow,\,\downarrow\label{equ:k1d-hybrid}
    \right\}
\end{equation}

\begin{equation}
    \mathcal{K}^{2d}\mathcal{A} = \left\{
    \alpha\int\mathrm{d}\bfr^\prime\mathcal{P}^c(\bfr,\bfr^\prime)\sum_{n^\prime}^{N_\mathrm{occ}^\sigma}\psi_{n^\prime\sigma}(\bfr^\prime)\int\mathrm{d}\bfr^{\prime\prime}v_c(\bfr^\prime,\bfr^{\prime\prime})a_{n^\prime\sigma}^*(\bfr^{\prime\prime})\psi_\nsigma(\bfr^{\prime\prime}):n=1,\cdots,N_\mathrm{occ}^\sigma;\sigma= \uparrow,\,\downarrow\label{equ:k2d-hybrid}
    \right\}
\end{equation}
where $f_\mathrm{Hxc}^\mathrm{loc} = v_c+f_\mathrm{xc}^\mathrm{loc}$ is the sum of the bare Coulomb potential and the local part of the exchange-correlation kernel
\begin{equation}
    f^\mathrm{loc}_\mathrm{xc}(\bfr,\bfr^\prime) = \frac{\delta V^\mathrm{loc}_\mathrm{xc}(\bfr)}{\delta\rho(\bfr^\prime)}
    \end{equation}
and the parameter $\alpha$ is the fraction of the Hartree-Fock exchange included in the definition of the hybrid functional. Note that the $\mathcal{K}^{1d}$ and the $\mathcal{K}^{2d}$ operators are zero for LDA/GGA functionals.

Once we have the solutions $a_\nsigma$ of the Liouville equation (Eq.~\eqref{equ:Liouville} or Eq.~\eqref{equ:general-stern}), i.e., the change of wavefunction $\partial\psi_\nsigma$, we can  compute the change of the density matrix with Eq.~\eqref{equ:change-of-gamma}; the change of density is then given by:
\begin{equation}
    \partial \rho(\bfr) = \partial \gamma(\bfr,\bfr) = \sum_{\sigma}\sum_{n}^{N_\mathrm{occ}^\sigma}\big[\partial\psi_\nsigma^*(\bfr)\psi_\nsigma(\bfr) + \psi_\nsigma^*(\bfr)\partial\psi_\nsigma(\bfr)\big]
\end{equation}
and force constants are obtained as follows:
\begin{equation}
    C_{I\alpha,J\beta}\propto\sum_\sigma\sum_n^{N_\mathrm{occ}^\sigma} \left\langle\partial_{I\alpha}\psi_\nsigma\left|\partial_{J\beta} V_\mathrm{ext}\right|\psi_\nsigma\right\rangle.
\end{equation}
By diagonalizing the dynamical matrix,
\begin{equation}
    \sum_{J\beta}\frac{1}{\sqrt{M_IM_J}} C_{I\alpha,J\beta} \xi_{J\beta,\nu} = \omega_{\nu}^2 \xi_{I\alpha,\nu},
\end{equation}
where $M_I$, $M_J$ are atomic masses, we obtain the frequency $\omega_\nu$ of mode $\nu$ and its  polarization $\xi_{I\alpha,\nu}$.

To compute the electron-phonon coupling matrices in the Cartesian basis:
\begin{equation}
    g_{m n I\alpha}^\sigma = \left\langle\psi_{m\sigma}\left|\partial_{I\alpha} V_\mathrm{scf}\right|\psi_\nsigma\right\rangle
\end{equation}
or in the phonon mode basis:
\begin{equation}
    g_{m n\nu}^\sigma =\sum_{I\alpha}\frac{\xi_{I\alpha,\nu}}{\sqrt{M_I}} g_{m n I\alpha}^\sigma,
\end{equation}
where $\xi_{I\alpha,\nu}$ is the $\nu$-th vibrational mode, we need to evaluate the change of the self-consistent (scf) potential $\partial V_\mathrm{scf}$. The scf potential is given by the sum of the Hartree potential $V_\mathrm{H}$, the local part of the exchange-correlation potential $V_\mathrm{xc}^\mathrm{loc}$ and the non-local Hatree-Fock exchange $V_\mathrm{xc}^\mathrm{nl}$. Thus, the change of the scf potential $\partial V_\mathrm{scf}\ket{\psi_{n\sigma}}$ is the sum of the following three terms:
\begin{equation}
    \partial V_\mathrm{H}(\bfr)\ket{\psi_
    \nsigma(\bfr)} = \psi_\nsigma(\bfr)\int\mathrm{d}\bfr^\prime v_c(\bfr,\bfr^\prime) \partial \rho(\bfr^\prime),
\end{equation}
\begin{equation}
    \partial V_\mathrm{xc}^\mathrm{loc}(\bfr)\ket{\psi_\nsigma(\bfr)} = \psi_\nsigma(\bfr)\int\mathrm{d}\bfr^\prime f_\mathrm{Hxc}^\mathrm{loc}(\bfr,\bfr^\prime)\partial \rho(\bfr^\prime),
\end{equation}
and
\begin{equation}
    \partial V_\mathrm{xc}^\mathrm{nl}\ket{\psi_\nsigma} = -\alpha\sum_{n^\prime}^{N_\mathrm{occ}^\sigma}\int\mathrm{d}\bfr^\prime\left[
        \partial\psi_{n^\prime\sigma}^*(\bfr^\prime)\psi_{n^\prime\sigma}(\bfr) + \psi_{n^\prime\sigma}^*(\bfr^\prime)\partial\psi_{n^\prime\sigma}(\bfr)
        \right]v_c(\bfr,\bfr^\prime) \psi_\nsigma(\bfr^\prime).
\end{equation}

The Fan-Migdal and Debye-Waller self-energies can then be computed as:
\begin{equation}
    \braket{\psi_\nsigma | \Sigma^\mathrm{FM}(\omega,T) |\psi_\nsigma } = \sum_{m\nu}\left|g_{m n\nu}^\sigma\right|^2\left[\frac{b_{\nu}(T)+f_{m\sigma}(T)}{\omega-\varepsilon_{m\sigma}+\omega_{\nu}-i0^+}+\frac{b_{\nu}(T)+1-f_{m\sigma}(T)}{\omega-\varepsilon_{m\sigma}-\omega_{\nu}-i0^+}\right]\label{equ:NA-FF-FM}
\end{equation}
\begin{equation}
    \braket{\psi_\nsigma | \Sigma^\mathrm{DW}(T) |\psi_\nsigma }=-\sum_{m\nu}\sum_{I\alpha J\beta}\frac{2b_{\nu}(T)+1}{\varepsilon_{n\sigma}-\varepsilon_{m\sigma}}\frac{1}{4\omega_{\nu}}\left[\frac{\xi_{I\alpha,\nu}\xi_{I\beta,\nu}}{M_I} +\frac{\xi_{J\alpha,\nu}\xi_{J\beta,\nu}}{M_J}\right]g^\sigma_{m n I\alpha}g^\sigma_{m n{J\beta}},
\end{equation}
where $b_\nu$ is the occupation number of the frequency $\omega_\nu$ obeying the Bose-Einstein distribution and $f_{m\sigma}$ is the occupation number of the Kohn-Sham eigenvalues $\varepsilon_{m\sigma}$ obeying the Fermi-Dirac distribution. The Debye-Waller self-energy is derived within the rigid-ion approximation (RIA) \cite{Allen1976,Ponce2014,Gonze2011}, which approximates second-order electron-phonon coupling matrices with first-order ones. 

Using the frequency-dependent Fan-Migdal self-energy, the renormalized energy levels can be evaluated self-consistently,
\begin{equation}
    \omega - \varepsilon_{\nsigma} = \braket{\psi_\nsigma | \Sigma^\mathrm{FM}(\omega,T)+\Sigma^\mathrm{DW}(T) |\psi_\nsigma }\label{equ:w-e=sigma}
\end{equation}
with initial guess $\omega_0=\varepsilon_\nsigma$, and using the Lanczos\cite{Lanczos1950} algorithm to evaluate the frequency-dependent Fan-Migdal self-energy (for a detailed derivation, see Ref.~\citenum{McAvoy2018} and Ref.~\citenum{Yang2021}). 

We refer to the FM self-energy in Eq.~\ref{equ:NA-FF-FM} as the non-adiabatic fully frequency-dependent (NA-FF) self-energy. If the frequency-dependence is considered within the adiabatic approximation, the self-energy is
\begin{equation}
    \braket{\psi_\nsigma | \Sigma^\mathrm{FM}_\mathrm{A-FF}(\omega,T) |\psi_\nsigma } \simeq \sum_{m\nu}|g^\sigma_{m n\nu}|^2 \frac{2b_{\nu}(T)+1}{\omega-\varepsilon_{m\sigma}}.
    \label{equ:FF-adiabatic-FM}
\end{equation}
We refer to Eq.~\eqref{equ:FF-adiabatic-FM} as the adiabatic fully frequency-dependent  (A-FF) self-energy.

In our formulation the evaluation of self-energies can be carried out simultaneously at multiple frequencies using the Lanczos algorithm; however, we introduce below approximations leading to frequency independent self-energies for comparison with results present in the literature, obtained e.g., with the Allen-Heine-Cardona (AHC) formalism\cite{Allen1976,Allen1981}. In particular, we evaluate the FM self-energy by applying the  so-called On-the-Mass-Shell (OMS) approximation, i.e., by setting $\omega=\varepsilon_{n\sigma}$ in the expressions of the A-FF and NA-FF  self-energies. In the former case we obtain the adiabatic AHC (A-AHC)\cite{Allen1976,Allen1981} approximation and in the latter case the non-adiabatic AHC (NA-AHC) approximation:
\begin{equation}
    \braket{\psi_\nsigma | \Sigma^\mathrm{FM}_\mathrm{A-AHC}(T) |\psi_\nsigma }\simeq\sum_{m\nu}\left|g^\sigma_{m n\nu}\right|^2\frac{2b_\nu(T)+1}{\varepsilon_{\nsigma}-\varepsilon_{m\sigma}}\label{equ:AHC-FM}
\end{equation}
\begin{equation}
    \braket{\psi_\nsigma | \Sigma^\mathrm{FM}_\mathrm{NA-AHC}(T) |\psi_\nsigma } = \sum_{m\nu}\left|g_{m n\nu}^\sigma\right|^2\left[\frac{b_{\nu}(T)+f_{m\sigma}(T)}{\varepsilon_\nsigma-\varepsilon_{m\sigma}+\omega_{\nu}-i0^+}+\frac{b_{\nu}+1-f_{m\sigma}}{\varepsilon_\nsigma-\varepsilon_{m\sigma}-\omega_{\nu}-i0^+}\right].\label{equ:non-adiabatic-AHC-FM} 
\end{equation}
We summarize the various levels of approximations applied to evaluate the FM self-energy in \autoref{tab:summary-of-FM-levels}; the corresponding DW self-energies are the same for all levels of approximation. Thus, we also use the acronyms A-AHC, NA-AHC, A-FF and NA-FF to denote the level of theory adopted  for the total self-energy (FM + DW) and for the electron-phonon renormalization of fundamental gaps. 

\begin{table}
    \centering
    \caption{A list of theoretical approximations used in this paper to compute the Fan-Migdal self-energy, where we specify  whether the on-the-mass-shell (OMS) and the adiabatic approximation are adopted ($\checkmark$) or not adopted ($\times$) }
    \begin{threeparttable}
    \begin{tabular}{cccc}
        \toprule[2pt]
        Level of theory                 & OMS          & Adiabatic    & Equation                                      \\
        \midrule
        (A-)AHC\tnote{a}                             & $\checkmark$ & $\checkmark$ & \eqref{equ:AHC-FM}              \\
        NA-AHC               & $\checkmark$ & $\times$     & \eqref{equ:non-adiabatic-AHC-FM} \\
        A-FF     & $\times$     & $\checkmark$ & \eqref{equ:FF-adiabatic-FM}           \\
        NA-FF & $\times$     & $\times$     & \eqref{equ:NA-FF-FM}                  \\
        \bottomrule[2pt]
    \end{tabular}
    \begin{tablenotes}
        \item[a] In the main text, we use AHC and A-AHC interchangeably.
    \end{tablenotes}
    \end{threeparttable}
    \label{tab:summary-of-FM-levels}
\end{table}



\section{Verification}\label{sec:verification}

To verify the implementation of the method described above in the \texttt{WEST}\cite{Govoni2015} package, we first computed the phonon frequencies of selected solids (diamond, silicon and silicon carbide) and the vibrational modes of selected molecules (\ce{H2}, \ce{N2}, \ce{H2O}, \ce{CO2}), and compared our results with  those of the  frozen phonon approach. In \autoref{tab:hybrid-elph:compare-solid-phonon} and \autoref{tab:hybrid-elph:compare-molecule-mode}, we summarize our results obtained at the PBE0\cite{Perdew1996PBE0}  level of theory and obtained by solving either the Liouville equation or by using the frozen phonon approach. The lattice constants used for diamond, silicon and silicon carbide are 
3.635, 5.464 and 4.372 \AA, respectively, and the cell used for molecules is a cube of edge $10.583$ \AA. For verification purposes,  we only computed the phonon modes at the $\Gamma$ point in the Brillouin zone of the solids. We used an energy cutoff of $60\,\mathrm{Ry}$ for the solids and $50\,\mathrm{Ry}$ for the molecules, and the SG15\cite{Schlipf2015} ONCV\cite{Hamann2013} pseudopotentials for all solids and molecules.

\autoref{tab:hybrid-elph:compare-solid-phonon} shows that the absolute difference of the phonon frequencies computed with the FPH approach and the method implemented here are small for silicon and silicon carbide, $0.19$ and $0.07\,\mathrm{cm^{-1}}$, respectively. The corresponding difference for diamond is larger, but still acceptable being below $5\,\mathrm{cm^{-1}}$.  In \autoref{tab:hybrid-elph:compare-molecule-mode}, we compare the vibrational frequencies of \ce{H2}, \ce{N2}, \ce{H2O} and \ce{CO2} molecules computed by solving the Liouville equation and applying the frozen-phonon approach. We found again that the differences are small for \ce{N2} and \ce{CO2}  (below $1\,\mathrm{cm^{-1}}$), albeit slightly larger for \ce{H2} and \ce{H2O}. The largest difference is found in the case of \ce{H2} ($17.30\,\mathrm{cm^{-1}}$), and this is most likely due to the numerical inaccuracy of the frozen-phonon approach.

\begin{table}
    \centering
    \caption{A comparison of selected phonon frequencies [$\mathrm{cm}^{-1}$] in diamond, silicon and silicon carbide  computed in a primitive cell with the PBE0 functional  by solving the Liouville's equation or by using the frozen-phonon approach.}
    \label{tab:hybrid-elph:compare-solid-phonon}
    \begin{tabular}{cccc}
        \toprule
        Solid           & Liouville & Frozen-phonon & Absolute difference \\
        \hline
        diamond         & 2136.21   & 2131.48       & 4.73                \\
        silicon         & 737.47    & 737.28        & 0.19                \\
        silicon carbide & 612.77    & 612.70        & 0.07                \\
        \bottomrule
    \end{tabular}
\end{table}

\begin{table}
    \centering
    \caption{A comparison of the vibrational modes [$\mathrm{cm}^{-1}$] of selected molecules obtained with the PBE0 functional and  computed by solving the Liouville's equation or by using the frozen-phonon approach. The symmetry of the mode is given in the second column.}
    \label{tab:hybrid-elph:compare-molecule-mode}
    \begin{tabular}{ccccc}
        \toprule
        Molecule & Symmetry & Liouville & Frozen-phonon & Absolute difference \\
        \hline
        \ce{H2}  & $a_1$    & 4421.48   & 4438.78       & 17.30               \\
        \ce{N2}  & $a_1$    & 2480.36   & 2480.36       & 0.00                \\
        \ce{H2O} & $a_1$    & 1652.79   & 1658.76       & 5.97                \\
        \ce{H2O} & $b_2$    & 3921.28   & 3936.57       & 15.29               \\
        \ce{H2O} & $a_1$    & 4033.68   & 4048.58       & 11.90               \\
        \ce{CO2} & $e_{1u}$ & 698.15    & 698.12        & 0.03                \\
        \ce{CO2} & $a_{1g}$ & 1375.10   & 1375.18       & 0.08                \\
        \ce{CO2} & $a_{1u}$ & 2419.08   & 2419.23       & 0.15                \\
        \bottomrule
    \end{tabular}
\end{table}

To verify our calculations of electron-phonon interactions, we carried out a detailed study of the renormalization of the HOMO-LUMO gaps ($E_g$) of the \ce{CO2}, \ce{Si2H6}, \ce{HCN}, \ce{HF} and \ce{N2} molecules, with the results for \ce{CO2} summarized in \autoref{tab:hybrid-elph:compare-co2-zpr} and the rest in \autoref{tab:hybrid-elph:compare-molecules-b3lyp}. 

\autoref{tab:hybrid-elph:compare-co2-zpr} summarizes the  renormalizations to the $E_g$ of the \ce{CO2} molecule obtained within the A-AHC formalism, and using  DFPT, FPH and path-integral molecular dynamics (PIMD)\cite{Kundu2021} at the LDA,\cite{Perdew1981} PBE,\cite{Perdew1996} PBE0\cite{Perdew1996PBE0} and B3LYP\cite{Becke1988,Becke1993,Lee1988} levels of theory, respectively.  With the LDA and PBE/GGA functionals, the solution of the Liouville equation  yields the same results as the method proposed in Ref.~\citenum{McAvoy2018} and Ref.~\citenum{Yang2021}, as expected.
When  solving the Liouville equation with the DFPT method, the rigid-ion approximation is adopted, however the latter approximation  is not used in the frozen-phonon approach, leading to a slight difference between the frozen-phonon and Liouville results. In addition, we carried out calculations with the hybrid functionals, PBE0 and B3LYP, and compared our results with those of the frozen-phonon and PIMD approaches\cite{Kundu2021}. The PIMD approach circumvents the rigid-ion approximation and also includes ionic anharmonic effects. 
Since the rigid-ion approximation is adopted and anharmonicity is not included in the Liouville approach, differences on the order of $\sim 30\,\mathrm{meV}$, relative to PIMD are considered as  acceptable. We note that the computed renormalizations of the gap of \ce{CO2} reported in the literature,\cite{Shang2021}   $-680.7$ and $-716.2\,\mathrm{meV}$ with LDA\cite{Perdew1981} and PBE+TS\cite{Tkatchenko2009} functionals, respectively, are significantly different from those obtained here. We also note  that Ref.\citenum{Shang2021} reports a result at the B3LYP level of theory, $-4091.6\,\mathrm{meV}$, which is  one order of magnitude larger than the corresponding LDA and PBE+TS results,  hence  calling into question the numerical accuracy of the data. Such significant differences between our and the results of Ref. \citenum{Shang2021}  probably stems from the different choices of basis functions, localized basis functions in Ref.~\citenum{Shang2021} and plane-waves in this work.

In addition to \ce{CO2}, we also computed the energy gap renormalizations of \ce{Si2H6}, \ce{HCN}, \ce{HF} and \ce{N2}  molecules with the B3LYP functional; these are shown in \autoref{tab:hybrid-elph:compare-molecules-b3lyp}. For \ce{Si2H6} and \ce{HCN}, the results computed with the Liouville's equation and the FPH approach agree well, with small differences of $22$ and $5\,\mathrm{meV}$, respectively. The renormalization of \ce{HF} is about $-20\,\mathrm{meV}$ with both the Liouville and FPH approaches, consistent with the result $-29.9\,\mathrm{meV}$ reported in literature. The Liouville and FPH methods both predict the renormalization of \ce{N2} to be close to zero, in agreement with Ref. \citenum{Shang2021}.

In summary, we have verified our implementation of phonon and electron-phonon interaction by comparing results computed with the Liouville's equation and those obtained with DFPT, FPH and PIMD methods. At the LDA/PBE level of theories, we obtain  exactly the same results as with DFPT, as expected;  at the hybrid functional level of theory, the results obtained with the Liouville's equation are comparable with those of the FPH and PIMD methods, with reasonable differences compatible with the different approximations employed in the three different approaches.

\begin{table}
    \centering
    \caption{Electron-phonon renormalization energies [$\mathrm{meV}$] of HOMO, LUMO energy levels and the HOMO-LUMO gap in the \ce{CO2} molecule, computed by solving the Liouville's equation, using density functional perturbation theory (DFPT), the frozen-phonon (FPH) approach and the path-integral molecular dynamics (PIMD) method. We compare results obtained with different functionals:   LDA, PBE, PBE0 and the B3LYP functionals,  and include results obtained in  Ref.~\citenum{Shang2021}. }
    \label{tab:hybrid-elph:compare-co2-zpr}
    \begin{tabular}{ccccc}
        \toprule
        Method                                      & Functional & HOMO Renorm. & LUMO Renorm. & Gap Renorm. \\
        \hline
        Liouville                                   & LDA        & 64           & -453         & -517        \\
        DFPT                                        & LDA        & 64           & -453         & -517        \\
        \hline
        Liouville                                   & PBE        & 65           & -350         & -415        \\
        DFPT                                        & PBE        & 65           & -350         & -415        \\
        FPH                                         & PBE        & 53           & -325         & -378        \\
        \hline
        Liouville                                   & PBE0       & 68           & -69          & -137        \\
        FPH                                         & PBE0       & 55           & -77          & -132        \\
        PIMD                                        & PBE0       & 59           & -103         & -162        \\
        \hline
        Liouville                                   & B3LYP      & 67           & -107         & -174        \\
        FPH                                         & B3LYP      & 54           & -89          & -143        \\
        PIMD                                        & B3LYP      & 58           & -112         & -170        \\
        \hline
        \multirow{3}{*}{Ref. \citenum{Shang2021}} & LDA        & ---          & ---          & -680.7      \\
                                                    & PBE+TS     & ---          & ---          & -716.2      \\
                                                    & B3LYP      & ---          & ---          & -4091.6     \\
        \bottomrule
    \end{tabular}
\end{table}

\begin{table}[htbp]
    \centering
    \begin{tabular}{cccc}
        \toprule
        Molecule   &    Liouville & FPH & Ref.~\citenum{Shang2021} \\
        \hline
        \ce{Si2H6} & -117 & -139  & -1872.3   \\
        \ce{HCN}   &  -19 &  -14 &  -171.4   \\
        \ce{HF}    &  -18 &  -25 &   -29.9   \\
        \ce{N2}    &    8 &   -6 &   8.7     \\
        \bottomrule
    \end{tabular}
    \caption{Electron-phonon renormalization energies [meV] of energy gap in \ce{Si2H6}, \ce{HCN}, \ce{HF} and \ce{N2} molecules computed by solving the Liouville's equation and using the frozon phonon (FPH) approach at the B3LYP level of theory.}
    \label{tab:hybrid-elph:compare-molecules-b3lyp}
\end{table}

\section{Electron-phonon renormalization of energy gaps in small molecules}\label{sec:small-molecule}
\begin{figure}
    \centering
    \includegraphics[width=1\textwidth]{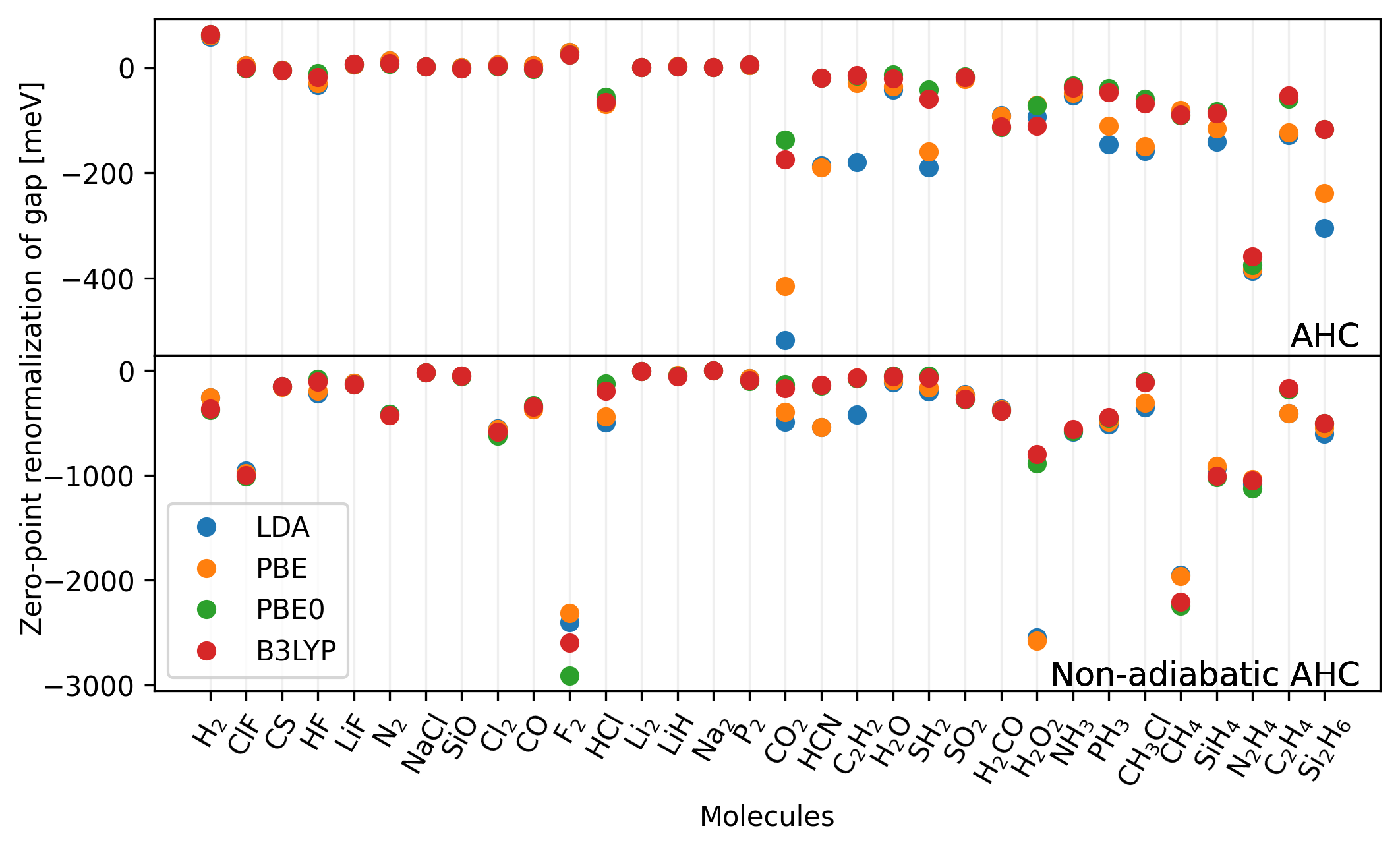}
    \caption{Computed zero-point renormalization energies of the HOMO-LUMO gaps of small molecules using the  AHC (upper panel) and NA-AHC (lower panel) approximations (see Table 1. for the definition of the approximations). We used different functionals specified in the inset.}
    \label{fig:hybrid-elph:molecule-zpr}
\end{figure}

Having verified our implementation, we carried out a study of the renormalization of the HOMO-LUMO gap of molecules in the G2/97 test set\cite{Curtiss1998} with LDA, PBE, PBE0 and B3LYP functionals. The results are summarized in \autoref{tab:hybrid-elph:molecule-adibatic-zpr} and \autoref{tab:hybrid-elph:molecule-nonadiabatic-zpr}, and are illustrated in \autoref{fig:hybrid-elph:molecule-zpr}.

\autoref{tab:hybrid-elph:molecule-adibatic-zpr} summarizes the renormalizations computed with the A-AHC formalism. For most of the molecules, using hybrid functionals does not significantly change the gap renormalization relative to LDA or PBE results. For example, the energy gap renormalizations of the \ce{H2} molecule computed with LDA, PBE, PBE0 and B3LYP functionals are $58\meV$, $61\meV$, $63\meV$ and $63\meV$, respectively. However, hybrid functionals do reduce the magnitude of gap renormalization in several systems, and \ce{CO2} and \ce{CH3Cl} are representative examples. In \ce{CO2} the renormalization is reduced from $-415\meV$ (PBE)  to $-137\meV$ (PBE0) level of theory; in \ce{CH3Cl}, it is reduced from $-149\meV$ (PBE) to $-59\meV$ (PBE0). 

We report in  \autoref{tab:hybrid-elph:molecule-nonadiabatic-zpr} our results within the non-adiabatic AHC (NA-AHC) framework. The removal of the adiabatic approximation significantly influences the computed magnitude of the gap renormalization in most of the molecules, with some exceptions, e.g. \ce{CO2}. For example, the \ce{H2} gap renormalization computed using PBE0 varies from $63\,\meV$ (AHC) to $-377\,\meV$ (NA-AHC). The most significant differences are found for the \ce{F2} and \ce{H2O2} molecules, where  the gap renormalizations computed at the PBE0 level of theory are $25$ and $-72\meV$, respectively, within the AHC approach and   $-2914$ and $-891\meV$, when using the non-adiabatic AHC method.

\begin{table}[htbp]
    \centering
    \caption{HOMO-LUMO energy gaps of small molecules and their zero-point renormalization energy (ZPR) computed within the adiabatic  Allen-Heine-Cardona (A-AHC) approximation. All gaps and ZPRs are in $\mathrm{eV}$. We compare results obtained with different energy functionals (LDA, PBE, PBE0 and B3LYP) and we also report ZPRs from Ref.~\citenum{Shang2021} and, in few cases, Ref.~\citenum{Gonze2011}}
    \label{tab:hybrid-elph:molecule-adibatic-zpr}
    \begin{threeparttable}

    \begin{tabular}{crrrrrrrrrr}
        \toprule
        \multirow{2}{*}{Molecule} & \multicolumn{3}{c}{LDA} & \multicolumn{2}{c}{PBE} & \multicolumn{2}{c}{PBE0} & \multicolumn{3}{c}{B3LYP}                                     \\
        \cmidrule(lr){2-4} \cmidrule(lr){5-6} \cmidrule(lr){7-8} \cmidrule(lr){9-11}
                         & gap    & ZPR      & Ref.~\citenum{Shang2021}     & gap     & ZPR       & gap    & ZPR    & gap    & ZPR     & Ref.~\citenum{Shang2021}    \\
        \hline
        \multirow{2}{*}{\ce{H2}}  & \multirow{2}{*}{9.998}  & \multirow{2}{*}{0.058}    &    -0.0021  & \multirow{2}{*}{10.164}  & \multirow{2}{*}{0.061}     & \multirow{2}{*}{11.890} & \multirow{2}{*}{0.063}  & \multirow{2}{*}{11.648} & \multirow{2}{*}{0.063}   &  \multirow{2}{*}{0.0036}\\
                                  &        &          &     0.0579\tnote{a}  &         &           &        &        &        &         &        \\
        \multirow{2}{*}{\ce{LiF}}         & \multirow{2}{*}{5.108}  & \multirow{2}{*}{0.006}    &     0.0331  & \multirow{2}{*}{4.723}   & \multirow{2}{*}{0.006}     & \multirow{2}{*}{7.014}  & \multirow{2}{*}{0.007}  & \multirow{2}{*}{6.601}  & \multirow{2}{*}{0.007}   &  \multirow{2}{*}{0.0040}\\
         &  &    &   0.0796\tnote{a}  &   &      &  &   &  &   &  \\
        \multirow{2}{*}{\ce{N2}}          & \multirow{2}{*}{8.221}  & \multirow{2}{*}{0.013}    &     0.0118  & \multirow{2}{*}{8.319}   & \multirow{2}{*}{0.013}     & \multirow{2}{*}{11.707} & \multirow{2}{*}{0.007}  & \multirow{2}{*}{11.179} & \multirow{2}{*}{0.008}   &  \multirow{2}{*}{0.0087}\\
                         &        &          &     0.0130\tnote{a}  &         &          &         &        &        &         &        \\
        \multirow{2}{*}{\ce{CO}}          & \multirow{2}{*}{6.956}  & \multirow{2}{*}{0.005}    &     0.0065  & \multirow{2}{*}{7.074}   & \multirow{2}{*}{0.004}     & \multirow{2}{*}{10.055} & \multirow{2}{*}{-0.003} & \multirow{2}{*}{9.575}  & \multirow{2}{*}{-0.002}  &  \multirow{2}{*}{0.0024}\\
                         &        &          &    0.0055\tnote{a}         &         &           &        &        &        &        & \\
        \ce{ClF}         & 3.194  & 0.004    &     0.0041  & 3.167   & 0.005     & 6.250  & -0.002 & 5.629  & -0.001  &  0.0025\\
        \ce{CS}          & 3.954  & -0.004   &    -0.0042  & 4.042   & -0.004    & 6.562  & -0.006 & 6.199  & -0.006  & -0.0058\\
        \ce{HF}          & 8.681  & -0.032   &    -0.0397  & 8.598   & -0.030    & 11.302 & -0.011 & 10.809 & -0.018  & -0.0299\\
        \ce{NaCl}        & 3.524  & 0.002    &     0.0001  & 3.225   & 0.002     & 5.069  & 0.002  & 4.577  & 0.002   &  0.0004\\
        \ce{SiO}         & 4.524  & 0.001    &    -0.0019  & 4.549   & 0.001     & 6.764  & -0.002 & 6.368  & -0.002  & -0.0032\\
        \ce{Cl2}         & 2.899  & 0.006    &     0.0063  & 2.894   & 0.006     & 5.503  & 0.002  & 4.887  & 0.003   &  0.0060\\
        \ce{F2}          & 3.495  & 0.030    &     0.0369  & 3.370   & 0.029     & 7.840  & 0.025  & 6.917  & 0.025   &  0.0329\\
        \ce{Li2}         & 1.532  & 0.001    &     0.0006  & 1.524   & 0.001     & 2.582  & 0.001  & 2.343  & 0.001   &  0.0007\\
        \ce{LiH}         & 2.985  & 0.002    &    -0.0066  & 2.873   & 0.003     & 4.424  & 0.001  & 4.117  & 0.002   & -0.0061\\
        \ce{Na2}         & 1.564  & 0.001    &     0.0002  & 1.521   & 0.001     & 2.495  & 0.000  & 2.264  & 0.001   &  0.0000\\
        \ce{P2}          & 3.649  & 0.005    &     0.0021  & 3.644   & 0.005     & 5.537  & 0.005  & 5.107  & 0.005   &  0.0037\\
        \ce{CO2}         & 8.075  & -0.517   &    -0.6807  & 8.033   & -0.415    & 10.159 & -0.137 & 9.708  & -0.174  & -4.0916\\
        \ce{HCN}         & 7.878  & -0.185   &    -0.1412  & 7.930   & -0.190    & 10.186 & -0.020 & 9.806  & -0.019  & -0.1714\\
        \ce{H2O}         & 6.272  & -0.042   &    -0.0806  & 6.208   & -0.036    & 8.511  & -0.013 & 8.084  & -0.020  & -0.0524\\
        \ce{SH2}         & 5.212  & -0.189   &    -0.0360  & 5.238   & -0.160    & 6.942  & -0.042 & 6.593  & -0.059  & -0.2117\\
        \ce{SO2}         & 3.457  & -0.019   &    -0.0178  & 3.414   & -0.021    & 6.087  & -0.016 & 5.596  & -0.018  & -0.0186\\
        \ce{H2CO}        & 3.470  & -0.091   &    -0.0876  & 3.589   & -0.092    & 6.451  & -0.114 & 5.993  & -0.111  & -0.1005\\
        \ce{H2O2}        & 5.028  & -0.093   &    -0.1290  & 4.887   & -0.071    & 7.780  & -0.072 & 7.505  & -0.110  & -0.2254\\
        \ce{NH3}         & 5.395  & -0.053   &    -0.0611  & 5.304   & -0.048    & 7.205  & -0.035 & 6.825  & -0.038  & -0.0333\\
        \ce{PH3}         & 5.999  & -0.146   &    -0.0592  & 5.946   & -0.110    & 7.388  & -0.039 & 7.056  & -0.047  & -0.2017\\
        \ce{C2H2}        & 6.703  & -0.179   &    -0.1901  & 6.712   & -0.029    & 8.181  & -0.016 & 7.835  & -0.014  & -0.2327\\
        \ce{CH3Cl}       & 6.232  & -0.158   &    -0.1441  & 6.210   & -0.149    & 8.042  & -0.059 & 7.691  & -0.068  & -0.1141\\
        \ce{CH4}         & 8.799  & -0.084   &    -0.1147  & 8.820   & -0.081    & 10.647 & -0.091 & 10.320 & -0.090  & -0.0947\\
        \ce{SiH4}        & 7.727  & -0.141   &    -0.6149  & 7.772   & -0.115    & 9.440  & -0.083 & 9.187  & -0.086  & -0.2027\\
        \ce{N2H4}        & 4.892  & -0.386   &    -0.1169  & 4.866   & -0.383    & 6.736  & -0.375 & 6.426  & -0.359  & -0.0793\\
        \ce{C2H4}        & 5.654  & -0.129   &    -0.1358  & 5.673   & -0.123    & 7.592  & -0.059 & 7.224  & -0.053  & -0.1194\\
        \ce{Si2H6}       & 6.364  & -0.305   &    -0.5880  & 6.386   & -0.238    & 7.874  & -0.117 & 7.609  & -0.117  & -1.8723\\
        \bottomrule
    \end{tabular}
    \begin{tablenotes}
        \item[a]  Ref.~\citenum{Gonze2011} 
    \end{tablenotes}
    \end{threeparttable}
\end{table}

\begin{table}[htbp]
    \centering
    \caption{HOMO-LUMO gaps of small molecules and their zero-point renormalization energies (ZPR) computed within the non-adiabatic Allen-Heine-Cardona (NA-AHC) approximation. All gaps and ZPRs are in $\mathrm{eV}$. We compare results obtained with different energy functionals (LDA, PBE, PBE0, B3LYP).}
    \label{tab:hybrid-elph:molecule-nonadiabatic-zpr}
    \begin{tabular}{ccrcrcrcr}
        \toprule
        \multirow{2}{*}{Molecule} & \multicolumn{2}{c}{LDA} & \multicolumn{2}{c}{PBE} & \multicolumn{2}{c}{PBE0} & \multicolumn{2}{c}{B3LYP}                                     \\
        \cmidrule(lr){2-3} \cmidrule(lr){4-5} \cmidrule(lr){6-7} \cmidrule(lr){8-9}
                                  & gap                     & ZPR                     & gap                      & ZPR                       & gap    & ZPR    & gap    & ZPR    \\
        \hline
        \ce{H2}                   & 9.998                   & -0.260                  & 10.164                   & -0.263                    & 11.890 & -0.377 & 11.648 & -0.366 \\
        \ce{LiF}                  & 5.108                   & -0.123                  & 4.723                    & -0.122                    & 7.014  & -0.134 & 6.601  & -0.134 \\
        \ce{N2}                   & 8.221                   & -0.418                  & 8.319                    & -0.432                    & 11.707 & -0.418 & 11.179 & -0.428 \\
        \ce{CO}                   & 6.956                   & -0.361                  & 7.074                    & -0.373                    & 10.055 & -0.338 & 9.575  & -0.346 \\
        \ce{ClF}                  & 3.194                   & -0.959                  & 3.167                    & -0.985                    & 6.250  & -1.011 & 5.629  & -1.000 \\
        \ce{CS}                   & 3.954                   & -0.151                  & 4.042                    & -0.156                    & 6.562  & -0.155 & 6.199  & -0.154 \\
        \ce{HF}                   & 8.681                   & -0.225                  & 8.598                    & -0.194                    & 11.302 & -0.083 & 10.809 & -0.111 \\
        \ce{NaCl}                 & 3.524                   & -0.021                  & 3.225                    & -0.022                    & 5.069  & -0.022 & 4.577  & -0.022 \\
        \ce{SiO}                  & 4.524                   & -0.052                  & 4.549                    & -0.054                    & 6.764  & -0.056 & 6.368  & -0.055 \\
        \ce{Cl2}                  & 2.899                   & -0.557                  & 2.894                    & -0.560                    & 5.503  & -0.622 & 4.887  & -0.589 \\
        \ce{F2}                   & 3.495                   & -2.405                  & 3.370                    & -2.317                    & 7.840  & -2.914 & 6.917  & -2.600 \\
        \ce{HCl}                  & 6.768                   & -0.501                  & 6.784                    & -0.440                    & 8.858  & -0.128 & 8.417  & -0.195 \\
        \ce{Li2}                  & 1.532                   & -0.007                  & 1.524                    & -0.008                    & 2.582  & -0.010 & 2.343  & -0.010 \\
        \ce{LiH}                  & 2.985                   & -0.049                  & 2.873                    & -0.045                    & 4.424  & -0.055 & 4.117  & -0.058 \\
        \ce{Na2}                  & 1.564                   & -0.002                  & 1.521                    & -0.002                    & 2.495  & -0.002 & 2.264  & -0.002 \\
        \ce{P2}                   & 3.649                   & -0.077                  & 3.644                    & -0.079                    & 5.537  & -0.100 & 5.107  & -0.096 \\
        \ce{CO2}                  & 8.075                   & -0.495                  & 8.033                    & -0.398                    & 10.159 & -0.136 & 9.708  & -0.174 \\
        \ce{HCN}                  & 7.878                   & -0.543                  & 7.930                    & -0.541                    & 10.186 & -0.147 & 9.806  & -0.138 \\
        \ce{H2O}                  & 6.272                   & -0.114                  & 6.208                    & -0.095                    & 8.511  & -0.050 & 8.084  & -0.061 \\
        \ce{SH2}                  & 5.212                   & -0.203                  & 5.238                    & -0.166                    & 6.942  & -0.050 & 6.593  & -0.069 \\
        \ce{SO2}                  & 3.457                   & -0.231                  & 3.414                    & -0.234                    & 6.087  & -0.281 & 5.596  & -0.274 \\
        \ce{H2CO}                 & 3.470                   & -0.364                  & 3.589                    & -0.376                    & 6.451  & -0.386 & 5.993  & -0.382 \\
        \ce{H2O2}                 & 5.028                   & -2.549                  & 4.887                    & -2.582                    & 7.780  & -0.891 & 7.505  & -0.799 \\
        \ce{NH3}                  & 5.395                   & -0.590                  & 5.304                    & -0.566                    & 7.205  & -0.578 & 6.825  & -0.562 \\
        \ce{PH3}                  & 5.999                   & -0.516                  & 5.946                    & -0.493                    & 7.388  & -0.453 & 7.056  & -0.450 \\
        \ce{C2H2}                 & 6.703                   & -0.420                  & 6.712                    & -0.074                    & 8.181  & -0.080 & 7.835  & -0.073 \\
        \ce{CH3Cl}                & 6.232                   & -0.351                  & 6.210                    & -0.307                    & 8.042  & -0.112 & 7.691  & -0.116 \\
        \ce{CH4}                  & 8.799                   & -1.950                  & 8.820                    & -1.961                    & 10.647 & -2.245 & 10.320 & -2.210 \\
        \ce{SiH4}                 & 7.727                   & -0.931                  & 7.772                    & -0.916                    & 9.440  & -1.019 & 9.187  & -1.007 \\
        \ce{N2H4}                 & 4.892                   & -1.082                  & 4.866                    & -1.038                    & 6.736  & -1.129 & 6.426  & -1.050 \\
        \ce{C2H4}                 & 5.654                   & -0.408                  & 5.673                    & -0.411                    & 7.592  & -0.184 & 7.224  & -0.173 \\
        \ce{Si2H6}                & 6.364                   & -0.607                  & 6.386                    & -0.551                    & 7.874  & -0.506 & 7.609  & -0.507 \\
        \bottomrule
    \end{tabular}
\end{table}

We emphasize that neither the AHC nor the non-adiabtic AHC formalism correctly describes the  self-energies in the full energy range, and thus we suggest that the frequency-dependent self-energies should always be computed.

\section{Electron-phonon renormalization of the energy gap of diamond}\label{sec:diamond}
We computed the electron-phonon renormalization of the energy gap in diamond within the AHC formalism, and by computing the NA-FF self-energies self-consistently (see \autoref{tab:summary-of-FM-levels} and Eq.~\eqref{equ:w-e=sigma}). The calculations for diamond were carried out in a $3\times 3\times 3$ supercell.

\begin{figure}
    \centering
    \includegraphics[width=\textwidth]{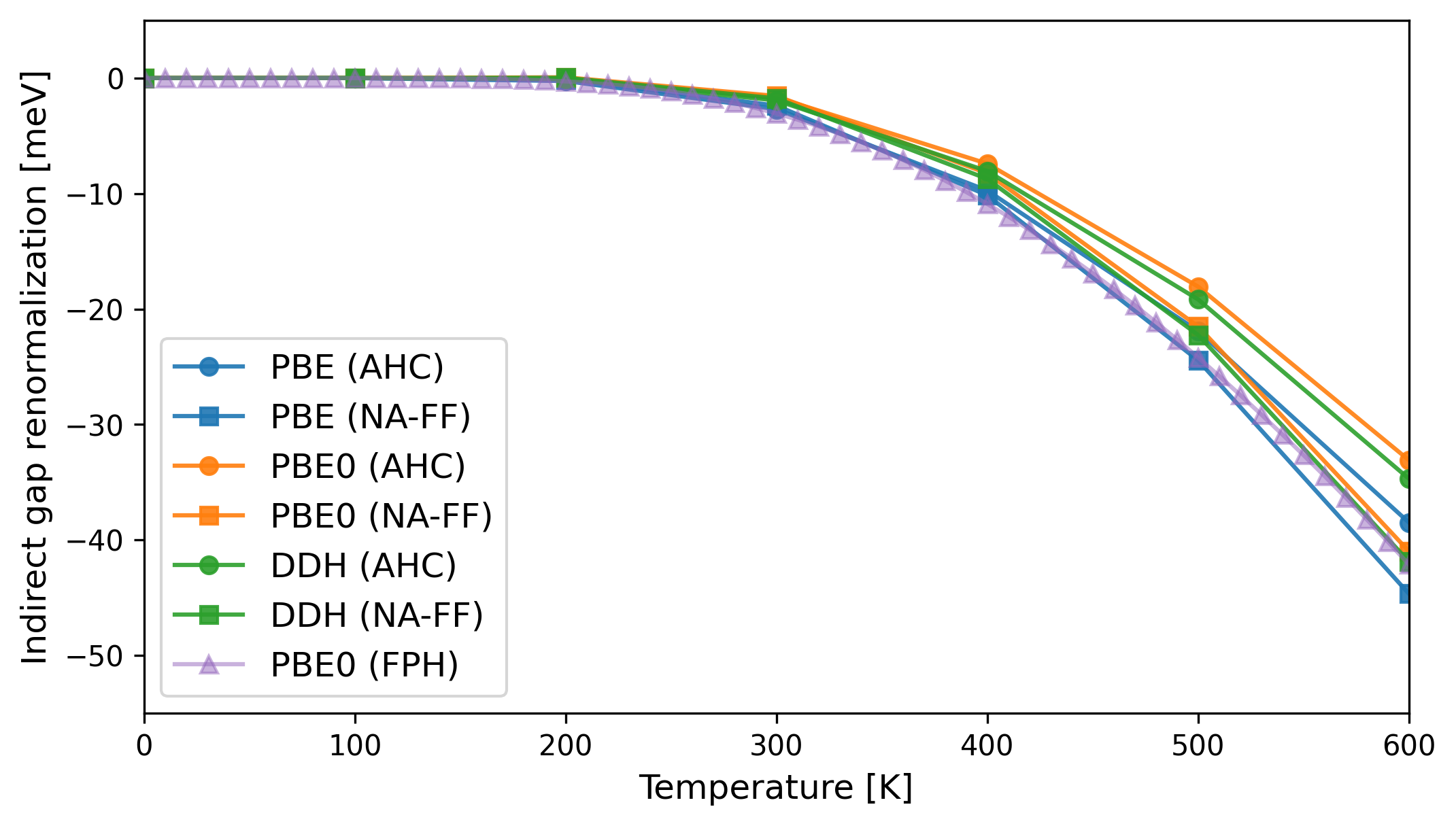}
    \caption{The electron-phonon renormalization energy of the indirect band gap of diamond computed by solving the Liouville equation and with different approximations to the self-energy, as defined in Table 1. The results obtained with the frozen phonon (FPH) approach and the PBE0 functional are also reported for comparison. The renormalization energy at zero temperature has been shifted to zero.}
    \label{fig:hybrid-elph:diamond-zpr}
\end{figure}

In \autoref{fig:hybrid-elph:diamond-zpr}, we present the temperature-dependent indirect gap renormalization computed with the PBE, PBE0 and dielectric dependent hybrid (DDH) functionals,\cite{Skone2014,Skone2016} where the fraction of exact exchange (0.18) in DDH is chosen to be the inverse of the dielectric constant of diamond (5.61).\cite{Skone2014} Within the same level of approximation, e.g., the AHC formalism (circles in the plot), the PBE, PBE0 and DDH results are almost the same for temperatures lower than $400\,\mathrm{K}$, but their difference increases at higher temperatures. With the same functional, e.g., the PBE0 functional (orange lines in the  plot), the results obtained with the fully frequency-dependent non-adiabatic self-energies are lower than those obtained with the AHC formalism. In general, the use of the hybrid functional does not significantly modify  the trend of the ZPRs computed at the PBE level, as a function of temperature.

In \autoref{fig:hybrid-elph:diamond-zpr}, we also report the renormalization of the indirect gap of diamond obtained with the frozen phonon approach and the PBE0 functional. The results obtained with the FPH (purple line) approach and the Liouville equation (orange lines in the plot) are essentiallyy the same below  $300\,\mathrm{K}$, but they differ as T is increased.  The difference between the AHC/NA-FF and FPH approaches is always smaller than $10\,\mathrm{meV}$ at all temperatures, and it is reasonable considering that the FPH approach does not adopt  the rigid-ion approximationg, which is instead used within the AHC and NA-FF approaches.

\begin{figure}
    \centering
    \includegraphics[width=\textwidth]{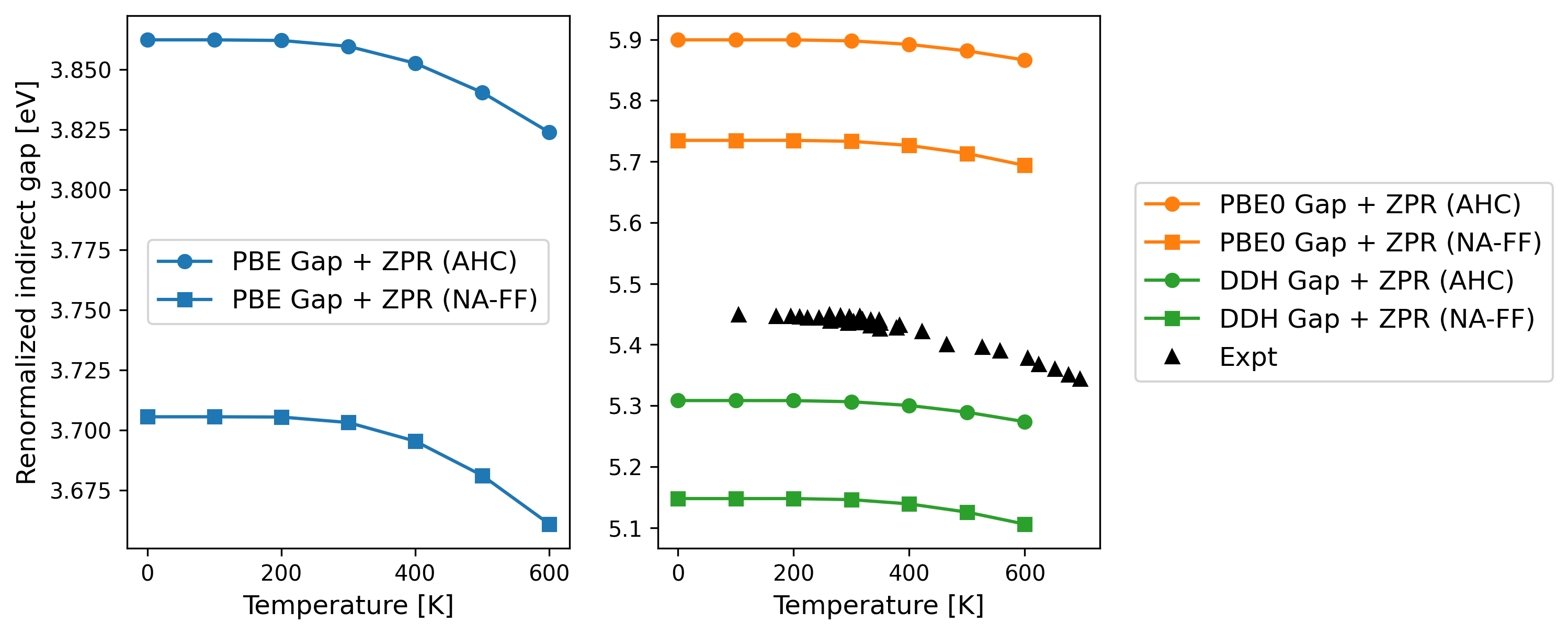}
    \caption{The electron-phonon renormalized indirect energy gap in diamond computed with the PBE and PBE0 functionals compared to experimental measurements.\cite{Odonnell1991} We show calculations performed with different approximations, as defined in Table 1.}
    \label{fig:hybrid-elph:diamond-renormalized-gap}
\end{figure}

\begin{table}
    \centering
    \caption{The temperature-dependent zero-point renormalization energy (ZPR) and renormalized indirect energy gap (Gap+ZPR) computed with the PBE, PBE0 and DDH functionals, using different levels of approximations, as defined in Table 1. The energy gaps computed at the PBE, PBE0 and DDH level of theory, without electron-phonon interaction, are $4.144$, $6.189$ and $5.597\,\mathrm{eV}$ respectively. All energies are reported in eV. }
    \label{tab:hybrid-elph:dimaond-renormalized-gap}
    \begin{tabular}{ccccccccccc}
        \toprule
                  & \multirow{2}{*}{Functional} & \multirow{2}{*}{Method} & \multicolumn{7}{c}{Temperature [K]} \\
        \cmidrule(lr){4-10}
        & & & 0 & 100 & 200 & 300 & 400 & 500 & 600 \\
        \cmidrule(lr){1-10}
        \multirow{6}{*}{ZPR} & \multirow{2}{*}{PBE} & AHC & -0.281 & -0.281 & -0.282 & -0.284 & -0.291 & -0.303 & -0.320 \\
         & & NA-FF & -0.438 &  -0.438 &  -0.438 &  -0.441 &  -0.448 &  -0.463 &  -0.483 \\
        \cmidrule(lr){2-10}
         & \multirow{2}{*}{PBE0} & AHC &-0.290 &  -0.290 &  -0.290 &  -0.291 &  -0.297 &  -0.308 &  -0.323 \\
         & & NA-FF & -0.454 &  -0.454 &  -0.454 &  -0.456 &  -0.463 &  -0.476 &  -0.495 \\
        \cmidrule(lr){2-10}
         & \multirow{2}{*}{DDH} &AHC& -0.289 &  -0.289 &  -0.289 &  -0.291 &  -0.297 &  -0.308 &  -0.324\\
         & &NA-FF& -0.450 &  -0.450 &  -0.450 &  -0.451 &  -0.458 &  -0.472 &  -0.492 \\
        \cmidrule(lr){1-10}
        \multirow{6}{*}{Gap+ZPR} & \multirow{2}{*}{PBE} & AHC & 3.862 &  3.862 &  3.862 &  3.860 &  3.853 &  3.840 &  3.824 \\
         & & NA-FF & 3.705 &  3.705 &  3.705 &  3.703 &  3.695 &  3.681 &  3.661 \\
        \cmidrule(lr){2-10}
         & \multirow{2}{*}{PBE0} & AHC &5.899 &  5.899 &  5.899 &  5.898 &  5.892 &  5.881 &  5.866\\
         & & NA-FF & 5.735 &  5.735 &  5.735 &  5.733 &  5.726 &  5.713 &  5.694 \\
        \cmidrule(lr){2-10}
         & \multirow{2}{*}{DDH} &AHC& 5.308 &  5.308 &  5.308 &  5.306 &  5.300 &  5.289 &  5.274 \\
         & &NA-FF& 5.148 &  5.148 &  5.148 &  5.146 &  5.139 &  5.125 &  5.106 \\
        \bottomrule
    \end{tabular}
\end{table}

A comparison of the  computed and measured renormalized energy gap of diamond is given in \autoref{fig:hybrid-elph:diamond-renormalized-gap} and  \autoref{tab:hybrid-elph:dimaond-renormalized-gap}. Although the PBE0 and DDH hybrid functionals yield a similar trend as PBE for the electron-phonon renormalization as a function of temperature, the renormalized gap are noticeably improved compared to experiments when using hybrid functionals. The indirect energy gap of diamond computed with PBE, PBE0 and DDH without electron-phonon renormalizaiton are $4.144$, $6.189$, and $5.597\,\mathrm{eV}$, respectively, and the experimental indirect gap measured  at approximately $100\,\mathrm{K}$ is $5.45\,\mathrm{eV}$.\cite{Odonnell1991}.  By including electron-phonon renormalizaiton, we can see that the results computed at the PBE0 level of theory agree relatively well with the experimental measurements (see \autoref{fig:hybrid-elph:diamond-renormalized-gap} and \autoref{tab:hybrid-elph:dimaond-renormalized-gap}). The renormalized indirect gap computed with the PBE0 functional at $100\,\mathrm{K}$ is $5.899\,\mathrm{eV}$ when the AHC formalism is used, and it is $5.735\,\mathrm{eV}$ when the  NA-FF self-energies are used. The renormalized indirect gaps computed with the DDH functional at $100\,\mathrm{K}$ are $5.308$ (AHC) and $5.148\,\mathrm{eV}$ (NA-FF).  As expected, the DDH results are closer to experimental measurements compared with those of the PBE0 functional, since the fraction of exact exchange is chosen according to the system specific dielectric constant. Overall we find that computing electron-phonon interactions at the hybrid level of theory is a promising protocol to obtain quantitative results, comparable to experiments.

\section{Application to spin defects in diamond}\label{sec:defect}
\begin{figure}[htbp]
    \centering
    \includegraphics[width=\textwidth]{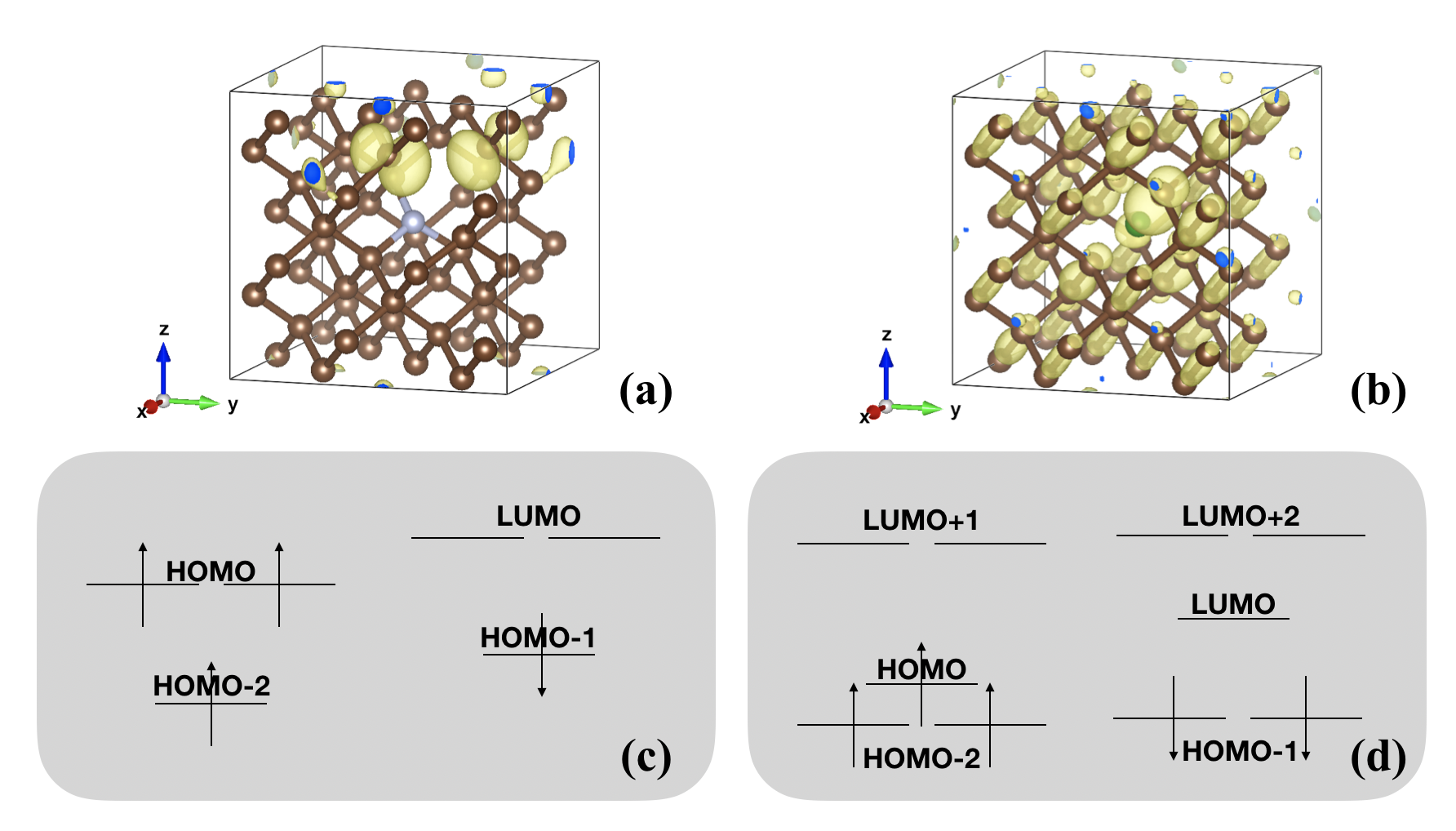}
    \caption{(a) The localized occupied state introduced by the nitrogen vacancy defect and (b) the delocalized unoccupied state introduced by the single boron vacancy defect. The wavefunctions are computed with the dielectric dependent hybrid functional. (c) and (d) illustrate the level ordering within the energy gap of diamond.}
    \label{fig:defect-states}
\end{figure}

Spin defects have been extensively studied due to their potential applications in quantum technologies.\cite{Weber2010,Jin2021,Cao2019,Huang2022} To accurately predict the electronic structures of spin defects, we computed their electronic properties using electron-phonon renormalizations and we considered a single Boron defect and the $\mathrm{NV}^-$ center shown in \autoref{fig:defect-states}. The calculations were carried out in a $2\times 2 \times 2$ cubic cell (63 atoms for $\mathrm{NV}^-$ center and 64 atoms for single Boron defect). 

In \autoref{tab:nv-center-zpr} and \autoref{tab:b-center-zpr}, we report the electronic energy levels and  zero-point renormalization for both defects obtained with the PBE, PBE0 and DDH functionals. We found that electron-phonon interactions weakly affect  the energy levels of the $\mathrm{NV}^-$ center, which exhibit localized wavefunctions; they are instead more significant for the single Boron defect with delocalized wavefunctions. In the $\mathrm{NV}^-$ center, the ZPR of the LUMO computed with the PBE functional is only $-35\,\mathrm{meV}$ and that of the HOMO is negligible. In addition, the hybrid functionals PBE0 and DDH yield results similar to PBE. For the boron defect,  with PBE (DDH) functional, the ZPRs of HOMO and LUMO are $111\,\mathrm{meV}$ (121) and $126\,\mathrm{meV}$ (241), respectively.

\begin{table}
    \centering
    \caption{Computed energy levels ($\mathrm{eV}$) and their zero-point renormalizations ($\mathrm{eV}$) in $\mathrm{NV}^-$ center. Energy levels are referred to the HOMO energy level, and the labels of energy levels are given in \autoref{fig:defect-states}(c).}
    \begin{tabular}{ccccccc}
        \toprule
             & \multicolumn{2}{c}{PBE} & \multicolumn{2}{c}{PBE0} & \multicolumn{2}{c}{DDH} \\
        \cmidrule(lr){2-3} \cmidrule(lr){4-5} \cmidrule(lr){6-7}
        & Level & ZPR & Level & ZPR & Level & ZPR \\
        \hline
        LUMO   &  1.359 & -0.035 &  3.593 & -0.033 &  2.948 & -0.033 \\
        HOMO   &  0.000 &  0.001 &  0.000 &  0.012 &  0.000 &  0.009 \\
        HOMO-1 & -0.411 &  0.012 & -0.059 &  0.004 & -0.189 &  0.008 \\
        HOMO-2 & -0.924 &  0.038 & -0.942 &  0.057 & -0.952 &  0.052 \\
        \bottomrule
    \end{tabular}
    \label{tab:nv-center-zpr}
\end{table}

\begin{table}
    \centering
    \caption{Computed energy levels ($\mathrm{eV}$) and their zero-point renormalizations ($\mathrm{eV}$) in  the boron defect. Energy levels are referred  to the HOMO energy level, and the labels of energy levels  are given in \autoref{fig:defect-states}(d).}
    \begin{tabular}{ccccccc}
        \toprule
             & \multicolumn{2}{c}{PBE} & \multicolumn{2}{c}{PBE0} & \multicolumn{2}{c}{DDH} \\
        \cmidrule(lr){2-3} \cmidrule(lr){4-5} \cmidrule(lr){6-7}
        & Level & ZPR & Level & ZPR & Level & ZPR \\
        \hline
        LUMO+2 &  4.061 & -0.359 &  6.109 & -0.367 &  5.517 & -0.365 \\
        LUMO+1 &  4.041 & -0.361 &  6.090 & -0.368 &  5.498 & -0.367 \\
        LUMO   &  0.137 &  0.126 &  1.389 &  0.285 &  1.027 &  0.241 \\
        HOMO   &  0.000 &  0.111 &  0.000 &  0.126 &  0.000 &  0.121 \\
        HOMO-1 & -0.278 &  0.087 & -0.319 &  0.054 & -0.308 &  0.062 \\
        HOMO-2 & -0.287 &  0.089 & -0.327 &  0.104 & -0.316 &  0.100 \\
        \bottomrule
    \end{tabular}   
    \label{tab:b-center-zpr}
\end{table}


\section{Conclusions}\label{sec:conclusion}
In this paper, we computed phonon frequencies and electron-phonon interaction at the level of hybrid density functional  theory by using density matrix perturbation theory and by solving the Liouville equation. Using this approach, we obtained phonon frequencies and energy gap renormalizations for molecules and solids by evaluating the non-adiabatic full frequency-dependent electron-phonon self-energies, thus circumventing the static and adiabatic approximations adopted in the AHC formalism, at no extra computational cost. We investigated the electronic properties of small molecules using LDA, PBE, B3LYP and PBE0 functionals. 
We also carried out calculations of the electronic structure of diamond with the PBE, PBE0 and DDH functionals, and found that the hybrid funtionals PBE0/DDH noticeably improve the renormalized energy gap compared to experimental measurements. In addition, we studied the electron-phonon renormalizations of defects in diamond, and we concluded that electron-phonon effects are essential to fully understand the electronic structures of defects, especially those with relatively delocalized states.

In conclusion, computing electron-phonon interactions at the hybrid functional level of theory is a promising protocol to accurately describe the electronic structure of molecules and solids, and density matrix perturbation theory is a general technique that allows one to do so in an efficient and accurate manner, by evaluating non adiabatic and full frequency dependent electron-phonon self-energies. 

\begin{acknowledgement}
This work was supported by the Midwest Integrated Center for Computational Materials (MICCoM) as part of the Computational Materials Sciences Program funded by the U.S. Department of Energy. This research used resources of the National Energy Research Scientific Computing Center (NERSC), a DOE Office of Science User Facility supported by the Office of Science of the U.S. Department of Energy under Contract No. DE-AC02-05CH11231, resources of the Argonne Leadership Computing Facility, which is a DOE Office of Science User Facility supported under Contract DE-AC02-06CH11357, and resources of the University of Chicago Research Computing Center.
\end{acknowledgement}


\bibliography{ref}

\end{document}